\documentclass[pra,twocolumn,showpacs,showkeys]{revtex4}

\usepackage{amssymb, amsmath, amsfonts, latexsym, verbatim, mathrsfs}

\usepackage{graphicx}

\begin{document}

\title{Minimum time generation of SU(2) transformations \\
with asymmetric bounds on the controls}
\thanks{Work supported by ARO MURI grant W911NF-11-1-0268}

\author{Raffaele Romano}
\email{rromano@iastate.edu}
\affiliation{Department of Mathematics, Iowa State University, Ames, IA (USA) }


\begin{abstract}

\noindent We study how to generate in minimum time special unitary transformations
for a two-level quantum system under the assumptions that: (i) the system is subject to a constant drift,
(ii) its dynamics can be affected by three independent, bounded controls, (iii) the bounds on the controls
are asymmetric, that is, the constraint on the control in the direction of the drift is independent of
that on the controls in the orthogonal plane. Using techniques recently developed for the analysis of $SU(2)$
transformations, we fully characterize the reachable sets of the system, and the optimal control strategies
for any possible target transformation.

\end{abstract}



\maketitle


\section{Introduction}

The implementation in minimum time of specific transformations is a key
ingredient of many protocols requiring the manipulation of two-level quantum systems (as
in quantum information theory~\cite{nielsen}, in quantum optics, or in atomic and molecular physics).
Usually, one is interested in mapping in minimum time an initial state to a final state under specific
conditions (see e.g.~\cite{boscain,carlini,hegerfeldt} and references therein). However, a general
and convenient approach to this problem consists in considering as control target the transformations themselves,
rather then the states of the underlying physical system. In our context, the problem can be
formulated as an optimal control problem on the Lie group $SU(2)$ of special unitary transformations in
$2$ dimensions~\cite{garon,raf,AD}. An arbitrary element of this group can be written as
\begin{equation}\label{su2}
    X = \left[
          \begin{array}{cc}
           \alpha & -\beta^* \\
            \beta & \alpha^* \\
          \end{array}
        \right],
\end{equation}
where $\alpha, \beta \in \mathbb{C}$ satisfy $\vert \alpha \vert^2 + \vert \beta \vert^2 = 1$.
The dynamics of $X$ is given by the Schr\"{o}dinger operator equation
\begin{equation}\label{evol}
    \dot{X} = (\omega_0 J_z + u_x J_x + u_y J_y + u_z J_z) X, \quad X(0) = I,
\end{equation}
where $J_k$ ($k = x, y, z$) are the skew-Hermitian generators of $SU(2)$, that is, independent elements
of the Lie algebra $\mathfrak{su} (2)$, $\omega_0$ is an arbitrary real parameter
characterizing a constant drift term in the dynamics, and $u_k = u_k (t)$ are possibly time-dependent
control actions constrained by
\begin{equation}\label{cons}
    u_x^2 + u_y^2 \leqslant \gamma_1^2, \quad u_z^2 \leqslant \gamma_2^2,
\end{equation}
where $\gamma_1 \geqslant 0$, $\gamma_2 \geqslant 0$. In other words, we assume that the strengths of the controls
affecting the dynamics through $J_z$, or rather through generators depending on $J_x$ and $J_y$, are independent.


Bounds on the controls depending on their squares, as in (\ref{cons}), naturally arise
when the control actions are physically realized through fields, with energy proportional to
their square amplitude. The choice of bounds in (\ref{cons}) corresponds to different
driving strengths along the $z$ direction, or along directions in an orthogonal plane.
This means that there is an anisotropy in the problem, with a privileged direction
in space, determined by the specific apparatuses which are used to steer the system.
Stronger constraints can be given by choosing independent bounds for the three controls,
$u_x^2 \leqslant \gamma_1^2$, $u_y^2 \leqslant \gamma_2^2$ and $u_z^2 \leqslant \gamma_3^2$,
and have been considered elsewhere (for instance, see \cite{boscain} and references therein).
From a theoretical point of view, the analysis of the specific constraints considered in this
work is relevant because it represents an intermediate situation between
the problem with independently constrained controls, which has not been solved in the general case,
and the problem with the isotropic bound $u_x^2 + u_y^2 + u_z^2 \leqslant \gamma^2$, which has been
fully investigated \cite{raf,AD}. Physically, bounds on the controls depending on their squares are
generically relevant in quantum information processing \cite{nielsen}, in atomic and molecular physics, and
in Nuclear Magnetic Resonance (NMR) \cite{levitt}.


The generators satisfy the standard $SU(2)$ commutation relations
\begin{equation}\label{comm}
    [J_j, J_k] = J_l,
\end{equation}
where $(j, k, l)$ is a cyclic permutation of $(x, y, z)$. More explicitly, $J_k = - \frac{i}{2} \sigma_k$,
where $\sigma_k$ are the Pauli matrices.

The target of the control action is to steer the identity $I = X (0)$ to an arbitrary final operator $X_f = X (t_f)$
in minimum time $t_f$, through a suitable optimal control strategy $u_k (t)$. To determine this strategy, we
will use the necessary condition of optimality provided by the {\it Pontryagin Maximum Principle} (PMP)~\cite{pontryagin,miko2},
which we briefly review. We introduce an auxiliary variable, the so-called {\it costate} $M \in \mathfrak{su}(2)$, represented
by the coefficients
\begin{equation}\label{cost}
    b_k = - \langle M, X^{\dagger} J_k X \rangle, \quad k = x, y, z,
\end{equation}
where $b_k = b_k (t)$, and $\langle A, B \rangle \equiv {\rm Tr} \, (A B^{\dagger})$. Then we define the
{\it Pontryagin Hamiltonian} as
\begin{equation}\label{pont}
    H(M, X, v_x, v_y, v_z) = \omega_0 b_z + v_x b_x + v_y b_y + v_z b_z.
\end{equation}
The PMP says that, if a control strategy $u_k(t)$ satisfying the bounds (\ref{cons}) and the corresponding trajectory $\tilde{X}(t)$
are optimal, then there exists a costate $\tilde{M} \ne 0$ such that
\begin{equation}\label{PMP}
    H(\tilde{M}, \tilde{X}, u_x, u_y, u_z) \geqslant H(\tilde{M}, \tilde{X}, v_x, v_y, v_z)
\end{equation}
for all $v_k$ satisfying (\ref{cons}). The PMP is only a necessary condition for optimality, useful
for finding extremal control strategies and trajectories. The optimal strategy and trajectory
are determined by comparing the extremal ones, analytically or numerically.

For its relevance, the control of $SU(2)$ operations has been extensively studied, with several constraints
on the control protocols. In this work we follow the approach presented in~\cite{raf}, which allows an
analytical investigation of the problem. In that paper, the cases with three or two controls were considered, with
bounds $u_x^2 + u_y^2 + u_z^2 \leqslant \gamma^2$ and $u_x^2 + u_y^2 \leqslant \gamma^2$
respectively. We refer to this work for several technical details which are omitted here for brevity.
Note that the present framework reduces to the case with two controls in~\cite{raf,AD} when $\gamma_2
= 0$. Moreover, some optimal solutions in the case with three controls
in~\cite{raf} are also optimal solutions in the present case, with values of $\gamma_1$ and $\gamma_2$
depending on the specific trajectory. Therefore, our analysis complements
that presented in~\cite{raf,AD}.

\section{Determination of extremal trajectories}

The differential equations describing the costate dynamics can be found  by using
(\ref{comm}) in (\ref{cost}), and they are given by
\begin{eqnarray}\label{bdyn}
  \dot{b}_x &=& - (\omega_0 + u_z) b_y + u_y b_z, \nonumber \\
  \dot{b}_y &=& (\omega_0 + u_z) b_x - u_x b_z, \\
  \dot{b}_z &=& u_x b_y - u_y b_x. \nonumber
\end{eqnarray}
After defining $\mu = \sqrt{b_x^2 + b_y^2}$, we find that $\mu^2 + b_z^2$ is a constant, which cannot vanish
because $M \ne 0$. By maximizing the Pontryagin Hamiltonian (\ref{pont}) with the constraints
(\ref{cons}), we determine the form of the extremal controls. There are three possible cases: (i) $\mu \equiv 0$;
(ii) $b_z \equiv 0$; (iii) neither of them. In case (i), it follows from (\ref{bdyn}) that $u_x \equiv 0$ and $u_y \equiv 0$,
and then $b_z$ is a non-zero constant. Therefore, it must be
\begin{equation}\label{optc1}
  u_z = \gamma_2 {\rm sign} (b_z).
\end{equation}
In case (ii), from maximization of $H$ we determine the form of the extremal controls
\begin{equation}\label{optc2}
  u_x = \gamma_1 \frac{b_x}{\mu}, \quad
  u_y = \gamma_1 \frac{b_y}{\mu},
\end{equation}
and $u_z$ has values in the interval $[-\gamma_2, \gamma_2]$.
Finally, in case (iii) both (\ref{optc1}) and (\ref{optc2}) must hold.

Having the form of the extremal controls, we can integrate (\ref{evol}) and determine the trajectories in $SU(2)$.
Case (i) is trivial, and we obtain $X(t) = e^{i (\omega_0 \pm \gamma_2) \tau} I$, where $\tau = \frac{t}{2}$.
We do not further consider this situation which does not provide optimal solutions. Cases (ii) and (iii)
can be jointly integrated, by remembering that $-\gamma_2 \leqslant u_z \leqslant \gamma_2$ or $u_z = \pm \gamma_2$
in the two cases, respectively. By using these controls in (\ref{bdyn}), it is possible to prove that
\begin{equation}\label{bdyn2}
  b_x = \mu \cos{(\omega t + \phi)}, \quad
  b_y = \mu \sin{(\omega t + \phi)},
\end{equation}
where $\phi$ is a constant, and $\omega$ is a possibly time-dependent function given by
\begin{equation}\label{omeg0}
    \omega = \omega (t) = \omega_0 + \frac{1}{t} \int_0^t u_z(s) \, ds
\end{equation}
in case (ii), and the constant
\begin{equation}\label{omeg}
    \omega = \omega_0 + u_z - \gamma_1 \frac{b_z}{\mu}
\end{equation}
in case (iii). Its range depends on the case under investigation: $\omega < \omega_0 + \gamma_2$
if $b_z > 0$,  $\omega > \omega_0 - \gamma_2$ if $b_z < 0$, and $\omega_0 - \gamma_2 \leqslant \omega (t) \leqslant
\omega_0 + \gamma_2$ if $b_z = 0$. We shall use $\omega$ rather than $b_z$ and $\mu$ to identify extremal trajectories.
Notice that we have been able to integrate (\ref{bdyn}) even in the case of time-dependent $u_z$ (and then $\omega$)
because of the simple form of this system. By substituting (\ref{bdyn2}) into (\ref{optc2}), we find
\begin{equation}\label{optcont2}
  u_x = \gamma_1 \cos{(\omega t + \phi)} ,\quad
  u_y = \gamma_1 \sin{(\omega t + \phi)}.
\end{equation}
By using these expressions in (\ref{evol}), and considering the representation of $X$
given in (\ref{su2}), following the steps detailed in \cite{raf} (which can be simply readapted
when $\omega$ is a function of time), we find
\begin{eqnarray}\label{solU}
     \alpha &=& e^{-i \omega \tau} \Big(\cos{a \tau} - i \frac{b}{a} \sin{a \tau}\Big) \nonumber \\
     \beta &=& -i \frac{\gamma_1}{a} e^{i (\omega \tau + \phi)} \sin{a \tau},
\end{eqnarray}
where we have rescaled time as $\tau = \frac{t}{2}$, and defined
\begin{equation}
  a = a (\omega) = \sqrt{b^2 + \gamma_1 ^2},
\end{equation}
with $b = 0$ in case (ii), $b = b (\omega) = \omega_0 + u_z - \omega$ in case (iii).

Because of the form of the drift term and the bounds on the controls, the problem has a natural cylindrical
symmetry. This is apparent from (\ref{solU}), where the phase of $\beta$ can be
arbitrarily modified through the parameter $\phi$. In other words, all operators $X$ differing
by an off-diagonal phase are completely equivalent in the framework adopted in this work, and they can be
reached in the same optimal time \cite{AD}. Consequently, we can fully describe
the extreme trajectories in $SU (2)$ by considering the evolution of $\alpha$, or, more conveniently,
its real and imaginary parts $x$ and $y$ respectively, which must satisfy $x^2 + y^2 \leqslant 1$.
Therefore, we can represent the relevant trajectories in the unit disk of $\mathbb{R}^2$ (or, equivalently,
in $\mathbb{C}$). When $b_z \ne 0$ they are given by
\begin{eqnarray}\label{xypm}
    x_{\pm} = \cos{\omega \tau} \cos{a \tau} -
  \frac{b}{a} \sin{\omega \tau} \sin{a \tau} \\
    y_{\pm} = - \sin{\omega \tau} \cos{a \tau} -
  \frac{b}{a} \cos{\omega \tau} \sin{a \tau}, \nonumber
\end{eqnarray}
with $\omega$ as in (\ref{omeg}); when $b_z = 0$ they are
\begin{equation}\label{xy0}
    x_0 = \cos{\omega \tau} \cos{\gamma_1 \tau}, \quad
    y_0 = - \sin{\omega \tau} \cos{\gamma_1 \tau},
\end{equation}\
with $\omega = \omega(t)$ as in (\ref{omeg0}).


In the analysis of extremal trajectories, it is often important to consider the so-called {\it singular}
trajectories, that is, extremal solutions such that the Pontryagin Hamiltonian is independent of the controls
(for the relevance of extremal trajectories in concrete problems, see for instance \cite{lapert,wu}
and references therein). By jointly considering (\ref{pont}) and (\ref{cost}), we can conclude that, in
the context considered in this work, these solutions do not exist, because they would require $b_x \equiv b_y
\equiv b_z \equiv 0$, which is inconsistent with the requirement $M \neq 0$. Only {\it regular} trajectories
(i.e., non singular) have to be taken into account. Using a similar argument, we observe
that it is impossible to concatenate the extremal trajectories described before. Again, this would require the
vanishing of all the $b_j$ at the switching time, which is not admitted.


\section{Evolution of the reachable sets}

Following~\cite{raf}, we define the
{\it optimal front-line} as the set of terminal points for a candidate
optimal trajectory at time $\tau$. As we have seen, depending on $b_z$, there are three families of
extremal trajectories. Correspondingly, there are three optimal front-lines in $\mathbb{R}^2$,
\begin{eqnarray}\label{newofl}
    {\cal F}_+ (\tau) &\equiv& \{ (x_+, y_+), -\infty < \omega < \omega_0 + \gamma_2 \}, \nonumber \\
    {\cal F}_- (\tau) &\equiv& \{ (x_-, y_-), \omega_0 - \gamma_2 < \omega < \infty \}, \\
    {\cal F}_0 (\tau) &\equiv& \{ (x_0, y_0), \omega_0 - \gamma_2 \leqslant \omega \leqslant \omega_0 + \gamma_2 \}, \nonumber
\end{eqnarray}
or similar definitions in $\mathbb{C}$, in terms of $\alpha_{0}$, $\alpha_{\pm}$. We remind that
$\omega$ is a constant for a given trajectory in ${\cal F}_+$ or ${\cal F}_-$, a possibly time-dependent
function for an extremal in ${\cal F}_0$.

The reachable set at time $\tau$ is, by definition, the set of operators in $SU(2)$ which can be
reached in time smaller or equal than $\tau$.
The evolution of the reachable set of the system is determined by the evolution of the optimal front
lines (\ref{newofl}), in particular, by their intersections, where the trajectories could lose
optimality. In the setting considered in~\cite{raf}, there is a unique optimal front line ${\cal F} (\tau)$,
and, if we work in $\mathbb{C}$ and forget for a while the bounds on $\omega$, ${\cal F}_{\pm} (\tau)$
can be expressed in terms of it as
\begin{equation}\label{funr2}
    {\cal F}_{\pm} (\tau) = e^{\mp i \gamma_2 \tau} {\cal F} (\tau)
\end{equation}
by means of suitable shifts in $\omega$. Therefore, assuming again that $\omega$ can be any real number, we can write
\begin{equation}\label{funr}
    {\cal F}_- (\tau) = e^{2 i \gamma_2 \tau} {\cal F}_+ (\tau),
\end{equation}
that is, at time $\tau$ the two sets are mapped into each other by a rotation of angle $2 \gamma_2 \tau$ in the unit disk.
We observe that ${\cal F}_+ (\tau) = {\cal F}_- (\tau)$ when $\tau = \frac{k \pi}{\gamma_2}$, with $k \in \mathbb{Z}$.

The individual analysis of ${\cal F}_+ (\tau)$ and ${\cal F}_- (\tau)$ follows from that of ${\cal F} (\tau)$.
We summarize the main results.
First of all, there is a one-to-one correspondence between values of $\omega$ and points on ${\cal F}_+$
and ${\cal F}_-$, that is, the associated trajectories do not intersect in optimal conditions.
Moreover, for each locus there is a critical trajectory~\footnote{When $\omega_0 = \pm \gamma_2$,
there is only one critical trajectory.} spiraling around the center of the disk,
modified with respect to that corresponding to ${\cal F} (\tau)$ according to (\ref{funr2}), and
parameterized by the critical frequencies
\begin{equation}\label{crifre1}
    \omega_c = \frac{\gamma_1^2 + (\omega_0 \pm \gamma_2)^2}{\omega_0 \pm \gamma_2},
\end{equation}
and losing optimality at the critical times
\begin{equation}\label{critime}
    t_c = \frac{\pi \vert \omega_0 \pm \gamma_2 \vert}{\gamma_1 \sqrt{(\omega_0 \pm \gamma_2)^2 + \gamma_1^2} }.
\end{equation}
These trajectories can be cut loci for the system, that is, special lines where optimal trajectories
lose their optimality. Other trajectories lose optimality on the border of the unit disk, which is, then, a
cut locus for the system. At time $\tau$, the frequencies corresponding to these trajectories are given by
\begin{equation}\label{crifre2}
    \omega_{c^{\prime}} (\tau) = (\omega_0 \pm \gamma_2) \pm \sqrt{\Big( \frac{\pi}{\tau} \Big)^2 - \gamma_1^2}.
\end{equation}
Note that, in (\ref{crifre1}) and (\ref{critime}), quantities with sign $+$ or $-$ refer to
${\cal F}_+$ or ${\cal F}_-$, respectively. The same applies for the first $\pm$ sign in (\ref{crifre2}),
but the second $\pm$ sign depends on the specific case. It is possible to prove that the possible scenarios are
$\omega_c > \omega_0 + \gamma_2 > 0$ or $\omega_c < \omega_0 + \gamma_2 < 0$ for ${\cal F}_+$, and
$\omega_c > \omega_0 - \gamma_2 > 0$ or $\omega_c < \omega_0 - \gamma_2 < 0$ for ${\cal F}_-$. Therefore, considering the allowed
range of values for $\omega$, we see that sometimes the critical trajectories are not extremal trajectories for the system.
We can also refine the definition of the optimal front lines, neglecting contributions which are certainly sub-optimal.
For instance, when $\omega_0 > 0$, the range of values of $\omega$ for ${\cal F}_+ (\tau)$ is given by $\omega_{c^{\prime}}
(\tau) < \omega < \omega_0 + \gamma_2$. For ${\cal F}_- (\tau)$, it is $\omega_{c} < \omega < \omega_{c^{\prime}} (\tau)$
when $\gamma_2 < \omega_0$, and $\omega_0 - \gamma_2 < \omega < \omega_{c^{\prime}} (\tau)$ when $\gamma_2 > \omega_0$.
Similar expressions can be found when $\omega_0 < 0$.

For small times (that is, in a neighborhood of $t = 0$)
there is a one-to-one correspondence between values of $\omega$ and points in ${\cal F}_0$.
Nonetheless, from (\ref{omeg0}) we see that there are different control strategies $u_z = u_z (t)$
leading to the same $\omega$, that is, those having the same time average.
Therefore, in this case there are different trajectories converging to the same
point of ${\cal F}_0$, and they are all equivalent \footnote{In fact, there are infinitely many trajectories
converging and departing from almost every point of ${\cal F}_0$, at any time. Exceptions are represented by
the trajectories corresponding to $\omega = \omega_0 \pm \gamma_2$, associated to only one control strategy,
with $u_z = \pm \gamma_2$ respectively.}.
In other words, in the region spanned by ${\cal F}_0$, there are distinct extremal trajectories (corresponding to different control
strategies) leading to the same final state in the same time, and remaining extremals after they intersect.
This is not in contradiction with the results of~\cite{raf}, where the optimal
solution is unique and $u_z$ is constant, because, in general,
the optimal solutions for the case of asymmetric bounds parameterized by $\gamma_1$ and $\gamma_2$
are not optimal solutions for the problem with symmetric bound given by $\gamma^2 = \gamma_1^2 + \gamma_2^2$.

Since $x_0^2 + y_0^2 = \cos^2{\gamma_1 \tau}$, the optimal front-line $\mathcal{F}_0 (\tau)$ is an arc of
circle centered at the origin, with time-dependent radius $\cos{\gamma_1 \tau}$, and angle $2 \gamma_2 \tau$.
If $2 \gamma_2 \tau \geqslant 2 \pi$, there are several values of $\omega$ corresponding to the same point on
the optimal front-line, and the corresponding extremal trajectories become equivalent, that is, they attain the
same point at the same time time even if the time average of $u_z (t)$ is different.

By considering the definition of $\mathcal{F}_0 (\tau)$ and $\mathcal{F}_{\pm} (\tau)$, we see that these
three loci are connected. Moreover, by using implicit differentiation, we find that
\begin{equation}\label{deriv}
    \frac{d y_0}{d x_0} = \frac{d y_{+}}{d x_{+}} = \frac{d y_{-}}{d x_{-}} =
    \cot{\omega \tau},
\end{equation}
therefore they are smoothly connected. It is possible to consider as optimal front line for this problem the
union of these three loci.

To complete the analysis, we must consider the intersections
between ${\cal F}_+ (\tau)$, ${\cal F}_- (\tau)$ and ${\cal F}_0 (\tau)$ at any time $\tau$. It turns out
that ${\cal F}_0 (\tau)$ never intersects  ${\cal F}_+ (\tau)$ or ${\cal F}_- (\tau)$ unless it is sub-optimal
(and then these intersections are irrelevant for the characterization of the evolution of the reachable sets
of the system). The intersection of  ${\cal F}_+ (\tau)$ and ${\cal F}_- (\tau)$ can be found numerically.
The two endpoints of ${\cal F}_+$ and ${\cal F}_-$, associated with $\omega = \omega_0 + \gamma_2$ and
$\omega = \omega_0 - \gamma_2$ respectively, coincide in two cases: either when $\tau = \frac{\pi}{\gamma_2}$
(when the two endpoints of ${\cal F}_0$ overlap), or when $\tau = \frac{\pi}{2 \gamma_1}$, when the radius of
${\cal F}_0$ vanishes.

\section{Typologies of evolution of the reachable sets and examples}

We can sum up the previous results, and classify
the systems in four classes, depending on the specific values of $\omega_0$, $\gamma_1$ and $\gamma_2$. They correspond
to different forms of the optimal trajectories, producing different time-evolutions of the reachable sets.
A given target operator $X_f \in SU(2)$ will require different control strategies (and associated minimum time
$t_f$) depending on the case at hand. For sake of simplicity we assume $\omega_0 \geqslant 0$, but a completely
analogous classification can be given also in the case $\omega_0 < 0$.

\begin{figure}[t]
  \includegraphics[width=8.5cm]{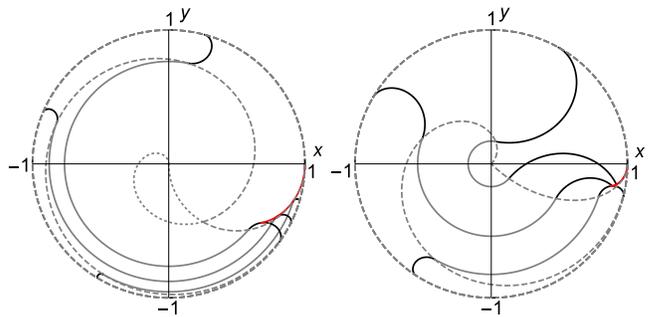}\\
  \caption{(Color online) Time evolution of the reachable sets in the unit disk, with $\omega_0 = 4$, $\gamma_1 = 1$ (left plot)
  or $2$ (right plot), $\gamma_2 = 3$. We have represented the optimal-front line at successive times $t = 0.6$, $1.0$ and $1.4$.
  The gray curve is $\mathcal{F}_0$, the dashed and dotted lines
  represent the evolution of its endpoints, before and after they converge, respectively.
  The region spanned by $\mathcal{F}_0$ and enclosed in the dashed lines contains points which can be reached
  via several equivalent optimal protocols (and, possibly, with different time average of $u_z (t)$ in the region enclosed
  in the dotted lines). The critical trajectory associated with $\mathcal{F}_-$ is a cut locus for
  the system. The other cut loci, not shown in the plot, are the border of the unit disk and the set of intersections
  between $\mathcal{F}_+$ and $\mathcal{F}_-$.}\label{Fig1}
\end{figure}

\begin{figure}[t]
  \includegraphics[width=8.5cm]{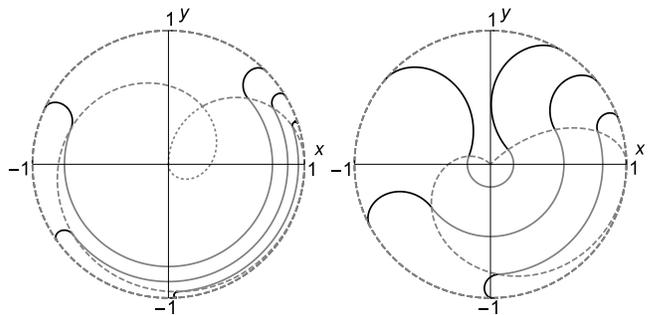}\\
  \caption{Time evolution of the reachable sets in the unit disk, with $\omega_0 = 2$, $\gamma_1 = 1$ (left plot)
  or $2$ (right plot), $\gamma_2 = 3$, at successive times $t = 0.6$, $1.0$ and $1.4$. The cut loci for
  the system, not shown in the plot, are the border of the unit disk and the set of intersections
  between $\mathcal{F}_+$ and $\mathcal{F}_-$.}\label{Fig2}
\end{figure}

First of all, we observe that, since $\omega_0 + \gamma_2 \geqslant 0$ the locus ${\cal F}_+$ rotates
counter-clockwise in the unit disk. The sense of rotation of ${\cal F}_-$ depends on the sign of
$\omega_0 - \gamma_2$, therefore the evolution of the reachable set is radically different in the two cases
$\gamma_2 > \omega_0$ and $\gamma_2 < \omega_0$. Similarly, the long-time evolution of the reachable
set depends on the relative magnitude of $\gamma_1$ and $\gamma_2$. Following the discussion of the
previous section, if $2 \gamma_1 \geqslant \gamma_2$ there are extremal trajectories with ending points on
${\cal F}_+$ and ${\cal F}_-$ which can get arbitrarily close to the center of the unit disk (corresponding
to the SWAP operator). Otherwise, when $2 \gamma_1 < \gamma_2$ there is a disk of radius $\cos{\pi
\frac{\gamma_1}{\gamma_2}}$, centered about the origin, whose points can only be reached by
trajectories associated with ${\cal F}_0$.
The relative magnitude between $\omega_0$ and $\gamma_1$ determines how many times the optimal
front lines spiral around the origin before exhibiting the aforementioned features.
For a graphical representations of the evolution of the reachable set with several choices of
the parameters see Fig.s \ref{Fig1} and \ref{Fig2}. In the first case we have $\gamma_2 > \omega_0$,
in the second case $\gamma_2 < \omega_0$, with two different settings for $2 \gamma_1$ and $\gamma_2$
in both cases.

Note that, if $\gamma_2 = 0$, the optimal front-line $\mathcal{F}_0$ collapses to a point and
the analysis consistently reproduces that of~\cite{raf} in the case of two independent controls.


To further illustrate the behavior of the reachable sets, we provide some examples of optimal synthesis
of some standard unitary gates, and compare with the corresponding results in the case of symmetric bounds.

In the case of
diagonal target operators, $X_f = e^{i \lambda \sigma_z}$ with $\lambda \in [0, 2 \pi)$,
since these are represented by points on the unit circle, which are reached by
the trajectories forming $\cal{F}_+$ or $\cal{F}_-$, the optimal control
strategies are given by controls $u_x$ and $u_y$ as in (\ref{optcont2}) with $\omega = \omega_{c^{\prime}}$
as in (\ref{crifre2}), and $u_z = \pm \gamma_2$, for $\cal{F}_\pm$ respectively. From (\ref{funr2}) and the results
in \cite{raf}, the optimal time is given by
\begin{equation}\label{optdiag}
    t_f = 2 \min_{u_z = \pm \gamma_2} \Big( \frac{(\pi - \lambda)(\omega_0 + u_z) + \Omega}{(\omega_0 + u_z)^2 + \gamma_1^2} \Big),
\end{equation}
where $\Omega = \sqrt{\pi^2 (\omega_0 + u_z)^2 + \lambda (2 \pi - \lambda) \gamma_1^2}$.
In particular, according to the former discussion on the evolution of the reachable sets, the minimum
is obtained with $u_z = \gamma_2$ when $\gamma_2 < \omega_0$; when
$\gamma_2 > \omega_0$ the situation is more complicated. The optimal time (\ref{optdiag}) and the corresponding
strategies can be compared with analogous time and strategies in the case of a symmetric bound on
the controls \cite{raf},
\begin{equation}\label{optidiag2ter}
    t_f = \left\{
            \begin{array}{ll}
              \frac{4 \pi - 2 \lambda}{\gamma + \omega_0}, & \hbox{if $\omega_0 \geqslant \frac{\pi - \lambda}{\pi} \gamma$} \\
              \frac{2 \lambda}{\gamma - \omega_0}, & \hbox{if $\omega_0 < \frac{\pi - \lambda}{\pi} \gamma$}
            \end{array}
          \right.
\end{equation}
obtained with $u_x = u_y = 0$ and $u_z = \gamma$ or $- \gamma$. If we require $\gamma^2 = \gamma_1^2 + \gamma_2^2$
(that is, the total control strength is the same), we have that the time (\ref{optidiag2ter}) is smaller than
(\ref{optdiag}), since the scenario with asymmetric control bounds is compatible with a symmetric bound.
As a check of consistency, this result can be proven by using the Lagrange multipliers method on (\ref{optdiag}), leading to
the constrained minimum (\ref{optidiag2ter}) obtained when $\gamma_1 = 0$ and $\gamma_2 = \gamma$.

As a second example, we choose as target operator the SWAP operator $X_f = i \sigma_y$, which represents the NOT
operation in quantum information. In this case, the behavior of the optimal trajectories is described by
the optimal front line $\cal{F}_0$, which we have analyzed in the previous section.
The optimal strategies are given by $u_x$ and $u_y$ as in (\ref{optcont2}) with $\omega$ taking an arbitrary
value in the interval $[\omega_0 - \gamma_2, \omega_0 + \gamma_2]$, possibly time-dependent. The control $u_z$
can take any form, in particular $u_z = 0$ can be chosen, leading to $\omega = \omega_0$.
For any choice of the parameters, the SWAP operator is attained in optimal time $t_f = \frac{\pi}{\gamma_1}$,
which is independent on $\gamma_2$, consistently with the fact that $u_z$ is completely irrelevant for
the optimal synthesis of this operator.
In the case of a symmetric bound on the controls, the optimal control strategy has a similar structure \cite{raf}, with
$u_x = \gamma \cos{(\omega_0 t + \varphi)}$, $u_y = \gamma \sin{(\omega_0 t + \varphi)}$ and $u_z = 0$ ($\varphi$ is
a phase), and the optimal time is $t_f = \frac{\pi}{\gamma}$. Again, this optimal time is smaller than that
with asymmetric bound on the controls, under the assumption that $\gamma^2 = \gamma_1^2 + \gamma_2^2$. It is clear that
the case with symmetric bound is reproduced when $\gamma_2 = 0$ and $\gamma_1 = \gamma$.


\section{Conclusions}

We have fully characterized the evolution of the reachable sets of system (\ref{evol})
with controls subject to the asymmetric constraints (\ref{cons}). We have derived the optimal trajectories
and the optimal control strategies for the system, and provided its classification in terms of the dynamical
parameters, which are completely arbitrary: driftless dynamics or unbounded control actions are special cases
of this treatment. By using this analysis, it is possible to compute, at least numerically, the minimum time for
generating an arbitrary $SU(2)$ transformation, and the required strategy.
For sake of clarity, we have analyzed the cases of diagonal operations, and the SWAP transformation
of a two-level system. These examples clearly illustrate the role of the constraints on the controls in the
study of time-optimal synthesis of quantum operations.


The main tool for studying the evolution of the reachable sets is the optimal front-line, which has already proven useful for the
investigation of the minimum-time synthesis of $SU(2)$ operations. Its application to similar
problems on different Lie groups is of great potential interest. It does not only provide a way to
clearly visualize the behavior of the reachable sets, but also a simple approach to prove rigorous results,
whose derivation when following the separate trajectories could be cumbersome in some regions of the
space of dynamical parameters.

For instance, the results in $SU(2)$
can be used to perform a similar analysis in $SO(3)$ (because of the standard homomorphism connecting
these groups) and are therefore related to the problem of attitude control of a rigid body. In this
context, asymmetric bounds on the controls, as those considered in this work, are especially relevant
 for the treatment of rigid bodies with rotational symmetry about one axis.


Another problem where the specific investigations presented in this paper could be of relevance is the synthesis
of $SU(2)$ operations with individual bounds on $u_x$, $u_y$ and $u_z$. Also in this case the
problem has more degrees of freedom, since the cylindrical symmetry of this work and \cite{AD,raf} is broken.
However, the analysis of suitable optimal front lines could provide new insights for the investigation of regions, in
space of dynamical parameters, which have not been studied so far.

Generalizations of this technique to problems characterized by an higher number of degrees of freedom
seems a promising research line. Despite in these cases it seems difficult to obtain the particularly simple
representation of the evolution of the reachable sets arising in $SU(2)$, we believe that
an approach based on the study of the envelopes of the front lines is more promising than a
direct analysis of the trajectories. A prospective direction for this line of research is the study
of the optimal implementation of two-qubit gates, or, more generally, the simultaneous control of
two spins (see \cite{sugny,raf2} for some recent applications of the PMP principle in this context).



\end{document}